\documentclass[12pt]{article}
\input{psfig.sty}

\def\Journal#1#2#3#4{{#1} {\bf #2}, #3 (#4).}
\def\NP{{\em Nucl. Phys.}}
\def\PLB{{\em Phys. Lett. B}}
\def\PR{{\em Phys. Rev.}}

\def\PRC{{\em Phys. Rev. C}}
\def\PRD{{\em Phys. Rev. D}}

\def\lp {\left( }
\def\rp {\right) }
\def\lb {\left[ }
\def\rb {\right] }
\def\lc {\left\{ }
\def\rc {\right\} }
\def\ra {\rangle }
\def\la {\langle }
\def\rar {\rightarrow }

\def\bb {\bibitem}
\def\nn {\nonumber}
\def\ni {\noindent}

\def\beq{\begin{equation}}
\def\eeq{\end{equation}}
\def\bea{\begin{eqnarray}}
\def\eea{\end{eqnarray}}

\def\a{\alpha}
\def\b{\beta}
\def\d {\delta}
\def\e{\epsilon}
\def\g{\gamma}
\def\D {\Delta}

\def\p {\pi}
\def\r{\rho}
\def\s{\sigma}

\def\x {x}
\def\id {d}
\def\rad {\eta}
\def\asym{\infty}

\def\bpi {\mbox{\boldmath $\pi$}}

\def\bnb {\mbox{\boldmath $\nabla$}}
\def\bp {\mbox{\boldmath $p$}}
\def\bP {\mbox{\boldmath $P$}}
\def\br {\mbox{\boldmath $r$}}
\def\bu {\mbox{\boldmath $u$}}
\def\bx {\mbox{\boldmath $y$}}
\def\by {\mbox{\boldmath $w$}}
\def\brm {r}
\def\bxm {y}
\def\bym {w}
\def\bz {\mbox{\boldmath $z$}}

\newcommand{\btau}{\mbox{\boldmath $\tau$}}

\newcommand{\bsig}{\mbox{\boldmath $\sigma$}}

\newcommand{\fpi}{f_\pi}

\def\Lg {{\cal L}}
\def\Ud {U^{\dagger}}
\def\hbpi {\hat{\bpi}}
\def\hbr {\hat{\br}}

\def\hbx {\hat{\bx}}
\def\hby {\hat{\by}}
\def\hbz {\hat{\bz}}

\def\cd {\!\cdot\!}
\def\dr {\partial }
\def\ub {\bar{u}}

\def\mpi {m_\pi}
\def\Tr {\mbox{Tr}}
\def\dim {\partial_{\mu}}
\def\dsm {\partial^{\mu}}
\def\din {\partial_{\nu}}

\date{}

\begin{document}

\title{Nucleon-nucleon interaction in the Skyrme model}

\author{Isabela P. CAVALCANTE 
\thanks{Present address: Instituto de F\'{\i}sica, Universidade do
Estado do Rio de Janeiro, Rio de Janeiro, RJ, Brazil. } 
\thanks{ipca@uerj.br}
\ and
Manoel R. ROBILOTTA \thanks{robilotta@if.usp.br}\\
Instituto de F\'\i sica, Universidade de S\~ao Paulo \\
C.P. 66318, 05315-970, S\~ao Paulo, SP, Brazil}

\maketitle

\begin{abstract}
We consider the interaction of two skyrmions in the framework of the sudden
approximation. The widely used  product ansatz is investigated. Its
failure in reproducing an attractive  central potential is associated
with terms that violate G-parity.
We discuss the construction of alternative ans\"atze  and identify  
a plausible solution to the problem.

\begin{noindent}\\
Keywords: Skyrme model, nucleon-nucleon interaction, chiral symmetry, 
effective Lagrangians.
\end{noindent}
\end{abstract}

\section{Introduction}
\label{s1}

Nucleon-nucleon $(NN)$ interactions are relatively simple at large
distances and become rapidly more complex as one moves inward.   
In the best phenomenological models existing at present, that
reproduce low-energy observables accurately, they are described 
by the consensual one-pion exchange potential (OPEP), supplemented by
theoretical two-pion exchange potentials (TPEP)  
and parametrized at short distances \cite{P,B}.  
The OPEP is responsible for a strong tensor component, which is mostly
important in few-body systems, such as the deuteron. 
Two-pion exchange, on the other hand, gives rise to the central
potential, that survives to all averages and is responsible for 
most properties of large systems and nuclear matter.
Quantum Chromodynamics (QCD) is the basic framework for the study of
strong processes and should have, in principle,  
an important role in the description of nuclear forces. 
However, at present, the non-Abelian character of this theory prevents
low energy calculations and one has to resort  
to effective theories, which must reflect the main features of QCD.
Thus, in Nuclear Physics applications, besides the usual space-time
invariances, one requires these theories to have approximate chiral
symmetry. 
The latter is usually restricted to the $SU(2)\times SU(2)$ sector,
for most processes involve only the quarks $u$ and $d$. 
This symmetry is explicitly broken by the small quark masses and, 
at the effective level, by the pion mass. 

Chiral symmetry has no influence over the OPEP, but is crucial to the 
TPEP, which depends on an intermediate pion-nucleon 
($\p N$) amplitude \cite{P}.
In the case of $NN$ interactions, the importance of this symmetry was
stressed already in the early seventies,  
by Brown and Durso \cite{BD71} and by Chemtob, Durso and Riska
\cite{CDR72}, who used it to constrain the form  
of the TPEP.
In that decade it also became popular to describe nuclear processes by
means of the linear $\s$ model \cite{GML},
containing a fictitious particle called $\s$ that, to some extent,
simulates the TPE. 
The elimination of this unobserved degree of freedom gave rise to
non-linear theories, which underlie modern descriptions  
of the interaction.
The first theoretical framework to incorporate non-linear chiral
dynamics into the $NN$ problem was proposed by Skyrme. 
This remarkable model for the nucleon, developed in the sixties
\cite{Sk} and revived in the eighties \cite{W}, describes baryons as
topological solitons, objects extended in space that rotate according
to the laws of Quantum Mechanics. 
The quark condensate appears as an intrinsic feature, corresponding to a
non-vanishing classical content of the vacuum, 
whose intensity is given by the pion decay constant $f_\p$. 
Skyrmions correspond to distortions of this condensate that
carry topological charges. 
One then works with pion fields which are unusually strong, in the
sense that their amplitudes may be comparable to $f_\p$. 
Thus, in spite of its well know limitations \cite{ad83}, the Skyrme model
remains a unique laboratory for studying chiral symmetry  
in the non-perturbative regime.

In the early nineties Weinberg restated the role of perturbative
chiral symmetry in nuclear interactions \cite{WNN} and motivated
interest in the TPEP.
Initially, several authors explored the pion-nucleon sector of
non-linear Lagrangians \cite{ChNN}, but the corresponding   
potentials could not reproduce even the medium range attraction in the
scalar channel. 
This happened because the TPEP is based on an intermediate $\p N$
amplitude, that can only be well described with the help  
of other degrees of freedom \cite{H83}.
Accordingly, in a later stage, agreement with empirical $\p N$
information was enforced and descriptions could reproduce $NN$
scattering data \cite{ORK,BRR,KBW}. 

In the case of perturbative calculations, the delta is by far the most
important non-nucleonic degree of freedom and is  
largely responsible for the intermediate range scalar attraction.
As the Skyrme model incorporates the delta from the very beginning,
one expects that it should yield a good qualitative $NN$
potential. 
However, it fails to do so.

Skyrme himself considered $NN$ interactions, already in the sixties,
using the so called product ansatz (PA) \cite{Sk}. 
The basic idea underlying the PA is that solutions corresponding to
baryon number $B=1$ can be used as building blocks  
to construct approximate solutions with an arbitrary value of $B$.
The great advantage of this approach is that the baryon number of the
composite system   is automatically equal to the number  
of individual $B=1$ skyrmions, irrespectively of their relative positions.
In the PA, the skyrmions that
constitute a larger system are assumed to retain their shape  
all the time, what is known as sudden approximation. 
In this framework, the construction of the $NN$ potential is rather
simple and  the fact that each nucleon has a  
profile function  which falls off rapidly with the distance allows one 
to assume that, for medium and large distances,
the $B=2$ system is not considerably different from the superposition
of two with $B=1$.
  
In the eighties, the product ansatz was used by Jackson {\em et al.}
\cite{ja85} and Vinh Mau {\em et al.} \cite{VM85}  
to calculate the $NN$ potential,  who found out a fully repulsive
central component, in disagreement with very well established
phenomenology. 
This puzzle motivated several attempts to construct improved versions
of those early works.  
Among them, one notes the symmetrized product ansatz by Nyman and
Riska \cite{nr88}, which could produce an intermediate range scalar
attraction.   
However, as pointed out by Sternheim and K\"albermann \cite{sk89},
there is a violation of baryon number conservation in this ansatz. 
Exact numerical calculations were also used, which allowed one to
evaluate the reliability of the sudden approximation at short
distances \cite{ww92}. 
Lattice calculations, using a method developed by Manton and
collaborators, gave rise to a torus-like baryon density,  
believed to correspond to the true $B=2$ ground state and having
almost twice the nucleon mass \cite{ver86}. 
The scalar potential associated with this configuration does show some
medium to long range attraction \cite{ww91}. 
However, it is worth recalling that lattice results depend on the
definitions adopted for collective coordinates and,  
as the full treatment is rather difficult, one usually resorts to
approximations \cite{ek96,gp98}.

In this work we consider the scalar interaction between skyrmions,  in
order to explore the possibility of obtaining the central attraction at large
distances by relaxing some of the constraints present
in usual calculations. 
We employ the sudden approximation because it gives rise to a
constructive  interaction, in which undeformed  nucleons are the main
building blocks, as in perturbative calculations.   
Our presentation is divided as follows.
In section \ref{s2} we study the asymptotic behaviour of the scalar
potential in the standard  product ansatz approximation,  
in order to understand why it does not yield attraction.
In section \ref{s3} we discuss the construction of alternative solutions,
which must be constrained to have the correct baryon number. 
Finally, in section \ref{s4} we analyze a possible
solution to the problem, and present concluding remarks.

\section{Central potential}
\label{s2}

The structure of the central potential has been studied recently, in
the framework of chiral perturbation theory 
\cite{KBW,pu00,R00}. 
In momentum space, the leading contribution has the generic form 
\beq
V_C(t) = -\;\frac{2}{\fpi^2 \mpi^2}\lb \fpi^2\lp a_{00}^+ + a_{01}^+
\, t \rp\rb \s(m;t) \,,
\label{cp1}
\eeq
where $\fpi$ is the pion decay constant, the $a_{0i}^+$ are
subthreshold coefficients \cite{H83} and $\s(m;t)$ is the scalar form
factor, that depends on both the momentum transferred $t$ and the
baryon  mass $m$.
This form factor is defined generically in terms of the symmetry breaking Lagrangian $\Lg_{sb}$ as
\beq
\la N(\bp')|- \Lg_{sb}\;|N(\bp) \ra = \s(t) \;\ub(\bp')\;u(\bp) \,.
\label{cp2}
\eeq 

In configuration space, eq.\  (\ref{cp1}) becomes 
\beq
V_C(\id) = -\;\frac{2}{\fpi^2 \mpi^2}\lb f_\p^2\lp a_{00}^+ + a_{01}^+
\bnb^2 \,\rp\rb \s(m;\id) \,,
\label{cp3}
\eeq
where $\id$ is the internucleon distance and $\s(m;\id)$ is the Fourier transform of $\s(m;t)$.
In order to allow this result to be compared with the corresponding one in the Skyrme model, we note that, in the large $N_c$
limit, the nucleon and the delta are degenerate and very heavy. 
In ref.\  \cite{pu00}, the heavy baryon limit of $\s(m;\id)$ was
considered and found to be
\beq
\s(m\rar\infty;\id) \rar \s^{HB}(\id) =
\frac{9\mpi^6}{128\p^2}\;\lp\frac{g_A}{\fpi}\rp^2  
\lb \frac{d}{d \x}  \;\frac{e^{-\x}}{\x}\rb^2\,,
\label{cp4}
\eeq
with $\x=\mpi \id$. The central potential is
\bea
V_C^{HB}(\id) &=& -\;\frac{2}{\fpi^2\mpi^2}\lb
\frac{9\mpi^6\;g_A^2}{128\p^2}\rb 
\lb a_{00}^+ \lp 1+\frac{2}{\x}+\frac{1}{\x^2}\rp\right.
\nn\\[2mm]
&+& \left. 4\;\mpi^2\;a_{01}^+ \lp
1+\frac{3}{\x}+\frac{11}{2\x^2}+\frac{6}{\x^3}+\frac{3}{\x^4}\rp \rb
\frac{e^{-2\x}}{\x^2} 
\label{cp5}
\eea
and, at very large distances, it behaves as 
\beq
V_C^{HB}(\id) \rar - K\; \lp \frac{e^{-\x}}{\x}\rp^2 \, .
\label{va}
\eeq
The sign of the constant $K$ is determined by the values of the
subthreshold coefficients in the combination  
$(a_{00}^+ +4\mpi^2 a_{01}^+)$.
In table \ref{t1} we display empirical values for the $a_{0i}^+$ and
it is possible to note that the correct sign of $V_C$ comes 
mainly from $a_{01}^+$, since $a_{00}^+$ in isolation would give rise
to a repulsive interaction. 

\vspace{2mm}

\begin{table}[h]
\begin{center}
\begin{tabular} {|c|c|c|c|c|}
\hline
& ref.\  \cite{H83}& ref.\  \cite{KH}& ref.\  \cite{PA}& ref.\  
\cite{PA} \\ \hline\hline
$ a_{00}^+\;\;( \mpi^{-1}) $ &$-1.46\pm0.10 $ &$-1.30\pm0.02$ &
$-1.27\pm0.03$ & $-1.15\pm0.03$ \\ \hline 
$ a_{01}^+\;\;( \mpi^{-3}) $ &$1.14\pm0.02$   &$1.35\pm0.14 $  &
$1.27\pm0.03 $  & $ 1.23\pm0.03$\\ \hline 
$ K $(MeV)  & 21.6 &28.5 & 26.5 & 26.2 \\ \hline 
\end{tabular}
\end{center}
\caption{Empirical values for the subthreshold coefficients
$a_{00}^+$, $a_{01}^+$ and the constant $K$, which  
determines the intensity of the central potential.}
\label{t1}
\end{table}

In order to study the central potential in the Skyrme model, we recall
that the standard soliton Lagrangian density is written as \cite{Sk,ad83}
\beq
\Lg = \Lg_\s + \Lg_4\,,
\label{lg1}
\eeq
where
\beq
\Lg_\s = \frac{\fpi^2}{4}\Tr  (\dim U\dsm \Ud) + \mpi^2\;
\frac{\fpi^2}{4}\; \Tr (U + \Ud -2) 
\label{lg2}
\eeq
\ni
corresponds to the non-linear $\s$ model and
\beq
\Lg_4 = \frac{1}{32e^2} \Tr \,[\dim U\Ud , \din U\Ud]^2
\label{lg3}
\eeq
\ni
is the stabilizing term. 
In these expressions, $e$ is a free parameter, called Skyrme constant,
whereas the dynamical variable $U$ is a  
$2 \times 2$ unitary matrix, given by
\beq 
U = e^{i \, \btau \cdot \hat {\bpi} F} =\cos F + i \, \btau \cd \hbpi \sin F,
\label{U}
\eeq
\ni
where $\btau$ are the isospin Pauli matrices and $F$ is the chiral
angle, whose boundary conditions determine  
the baryon number of a particular configuration.
The function $F$ and the isospin direction $\hbpi$ are related to the
pion field $\bpi$ of the non-linear $\s$ model \cite{GML} by
$
\bpi = \fpi  \sin F \, \hat{\bpi}.
$

In the $B=1$ case, a static solution is obtained using the condition
$\hbpi = \hbr$, the so called hedgehog ansatz, with boundary
conditions $F(r=0) = \pi$ and $F (r\rar\infty) = 0$ \cite{ad83}.  
The quantization of  this baryon is achieved by rotating the static
solution with the help of collective coordinates, as a rigid body.  
This procedure endows the skyrmion with spin and isospin and
corresponds to multiplying the pion field by the rigid  
body rotation matrix $D$,
\beq
\pi_i \rar \pi_\a^q = D_{\a i}\; \pi_i.
\label{q}
\eeq
The matrices $D$ satisfy the completeness relations 
$D_{\a i}D_{\a j}= \d_{ij} , \; D_{\a i}D_{\b i}= \d_{\a \b}$ 
and, in the case of  nucleons, the correspondence with the ordinary
formalism is achieved by using 
\beq
\la N|D_{\a i} |N\ra  = -\; \frac 1 3  \la N|\tau_\a\; \s_i |N\ra \, ,
\label{D2}
\eeq
$\s_i$ being the spin Pauli matrices.

The scalar form factor in the Skyrme model can be obtained directly
from eqs.\ (\ref{lg2}) and (\ref{U}) and reads 
\beq
\s^{Sk}(\id) = \la N| -\Lg_{sb}(\id)\;| N\ra = - \mpi^2 \fpi^2 \lb
\cos F(\id) -1\rb \, .
\label{sSk}
\eeq

On the other hand, the asymptotic form of the chiral angle is
determined by $\Lg_\s$ as \cite{ad87}
\beq
F_\asym (\id) = \lp \frac{3 g_A \mpi^2}{8\pi\fpi^2}\rp \lp \frac{d}{d
\x}\;\frac{e^{-\x}}{\x}\rp 
\label{fa}
\eeq
and hence, for large distances, the leading term in eq.\ (\ref{sSk})
yields $\s_\asym^{Sk}(\id) = \s^{HB}(\id) $.
In order to test this relationship further, we write 
\bea
\s_\asym^{Sk}(\id) &=& \mpi^2 \fpi\; \la N| \sqrt{\fpi^2-\bpi^q\cd\bpi^q}-\fpi |N\ra 
\;\simeq \frac{\mpi^2}{2}\;  \bpi^2 
\nn\\[2mm]
&=& \frac{\mpi^2}{2}\lb \la N| \pi_\a^q | N\ra\la N| \pi_\a^q | N\ra + \la N| \pi_\a^q | \D\ra\la \D| \pi_\a^q | N\ra \rb\,.
\label{sSka}
\eea

Using eqs.\  (\ref{q}) and (\ref{D2}) in the last expression, one concludes that $N$ and $\D$ intermediate states 
determine respectively 1/3 and 2/3 of the total value of $\s_\asym^{Sk}(\id)$.
This relative proportion is identical to that found recently in the framework of chiral perturbation theory \cite{R00}, 
indicating that the Skyrme model does provide a rather reliable description of the scalar form factor in the heavy baryon limit.

For systems with $B=2$, the standard point of departure for constructing approximate solutions is the product ansatz (PA).
It uses two undistorted $B=1$ hedgehog solutions, whose centers 
are located at two fixed points  equidistant from origin along the $z$ 
axis, so that the hedgehog space
 coordinates are given by 
$\bx = \br +  \id \, \hbz /2$ 
and $\by = \br - \id \, \hbz /2 $.
Denoting the composite field by $U (\bx, \by)$, one writes
\beq
U (\bx, \by) = U (\bx) \, U (\by).
\label{pa}
\eeq

In this configuration, the $B=2$ condition is automatically fulfilled,
for any distance between their centers \cite{za86}. 
As the PA keeps the identities of constituent skyrmions, it allows the
direct incorporation of spin and isospin, through collective
rotations of individual hedgehogs. 

The potential is a function of the distance $\id$ and given by 
\beq
V(\id) = - \int d^3r \, \Lg_{int}(\br,\id\hbz) \;, 
\label{pot1}
\eeq
where $\Lg_{int}$ is obtained by using the field $U(\bx,\by)$ in the Skyrme Lagrangian, eqs.\ (\ref{lg1}-\ref{lg3}),  and  
subtracting the self energies $\Lg [U(\bx)]$ and $\Lg [U(\by)]$.
This potential works well in the isospin-dependent channels, since the OPEP is reproduced for distances larger than $2$~fm
and it is also possible to identify the roles of $\r$ and $A1$ mesons \cite{ya85}.
On the other hand, problems occur in the scalar-isoscalar channel, where the interaction is repulsive at all distances, in sharp contradiction with phenomenology. 

Using the definitions $F'_\brm= dF (\brm)/ d\brm$, $s_\brm=\sin
F(\brm)$ and $c_\brm =\cos F(\brm)$, the central potential is given by
\bea
V_C^{pa}(\id) &=& \frac {2 \fpi}{e} \frac{4 \pi}{3} \int_0^\infty dz \int_0^\infty \r \; d \r 
\lc -\; \frac {3 \mpi^2}{16 e^2 \fpi^2}\; (1-c_\bxm)(1-c_\bym) \right. 
\label{vc-form}  \\[3mm]
&+& \left. \lb \lp F'^2_\bxm + \frac {s_\bxm^2}{\bxm^2}\rp \lp F'^2_\bym + \frac{s_\bym^2}{\bym^2}\rp 
+  \frac {2 s_\bxm^2 s_\bym^2}{\bxm^2 \bym^2}- \lp \hbx \cd \hby \rp^2 
\lp F'^2_\bxm - \frac {s_\bxm^2}{\bxm^2}\rp \lp F'^2_\bym - \frac {s_\bym^2}{\bym^2}\rp \rb \rc. 
\nonumber
\eea

In order to study its asymptotic structure, we note that the pion
fields exist effectively only in the neighbourhood  
of the hedgehog centers. 
When the distance $\id$ is large the skyrmion located at
$(0,0,\id/2)$ is in the presence of the asymptotic region of
$U(\bx)$, 
we expand $F_\bxm$, $F'_\bxm$ and $\hbx\cd\hby$ around the point
$\by=0$ and write 
\bea
F_\bxm & \simeq & \a \; e^{-\mpi \bym_z} \lp 1 + \frac {f_1}{\x} +
\frac {f_2}{\x^2}\rp \frac {e^{-\x}}{\x}\;,  
\label{fx}\\[2mm]
F'_\bxm &\simeq& - \a \; e^{-\mpi \bym_z}\lp 1 +  \frac {g_1}{\x} +
\frac {g_2}{\x^2}\rp \frac {e^{-\x}}{\x} \;, 
\label{fpx}\\[2mm]
\hbx \cd \hby &\simeq& \frac {\bym_z}{\sqrt{\r^2 + \bym_z^2}} \lp 1 +
\frac {\mpi \r^2}{\bym_z \x}- \frac {3\mpi^2 \r^2}{2 \x^2} \rp\;, 
\label{xy}
\eea 
where $f_i$, $g_i$ are dimensionless polynomials of $\bym_z\equiv
(z-\id/2)$ and $\rho$, which are not displayed here. 
These expressions were tested order by order, by using them in
eq.\ (\ref{vc-form}) and checking that the potential did had 
the asymptotic structure, as in eq.\ (\ref{va}). 
We found out that it was necessary to expand $F(\bx)$ up to order
$\id^{-2}$, in order to have accurate  results.

Replacing eqs.\ (\ref{fx}-\ref{xy}) into  (\ref{vc-form}), we obtain an asymptotic contribution of the form
\beq
V_C^{pa} (d) \rar -K \lb 1 + \frac {\a_1} {\x } + \frac {\a_2}{\x^2} \rb \frac {e^{-2\x}}{\x^2}\;,
\label{asym-form}
\eeq
for both  $\Lg_\s$ and $\Lg_4$, separately. 
The values of the parameters $K$ and $\a_i$  
are displayed in table \ref{tabmulti},
based on the numerical constants $\mpi = 139$~MeV, $\fpi = 93$~MeV and $e=4.0$.
For the sake of comparison, we also present the values of those
parameters in the case of the phenomenological Argonne potential \cite{ar84}.

\begin{table}[h]
\begin{center}
\begin{tabular}{|c|c|c|c|c|}
\hline
& ${\cal L}_\sigma$ & ${\cal L}_4$ & ${\cal L}_\sigma$+${\cal L}_4$ &
Argonne \\ 
\hline\hline
$ K$(MeV) & 7.29  & -8.52  &  -1.23  & 4.80 \\ \hline
$\a_1$ &  0.93 &  4.31 &  24.4 & 1.0 \\ \hline
$\a_2$ & 1.78 & 12.9 & 79.1 & 6.0 \\ \hline
\end{tabular}
\end{center}
\caption{Coefficients of the multipole expansion of $V_C$ for the
product ansatz and the Argonne potential, as defined
in eq.\ (\ref{asym-form}).}
\label{tabmulti}
\end{table}

\begin{figure}[hbt]
\centerline{\psfig{figure=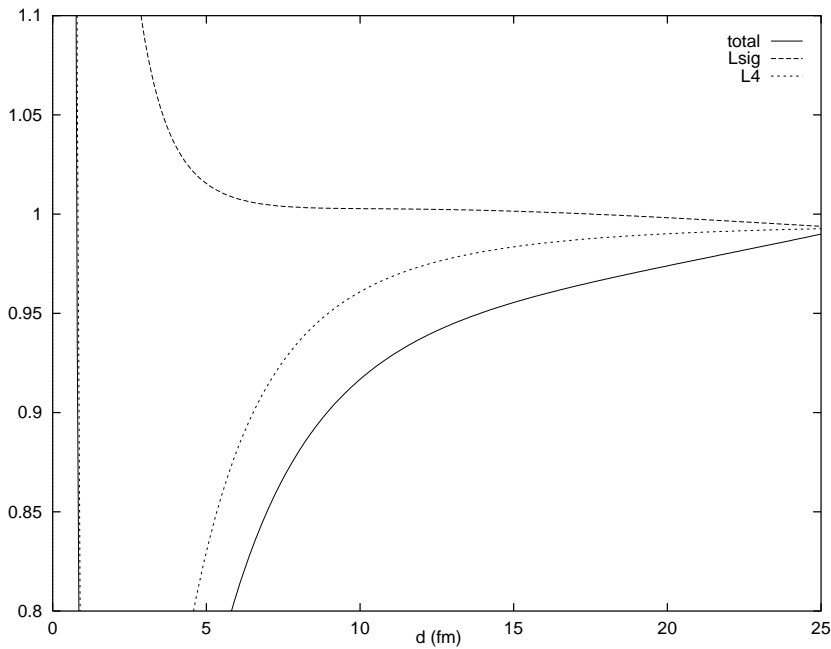,width=12cm}} 
\vspace*{0.3cm}
\caption{Ratios between  the multipole expansion and the exact
numerical result for the scalar potential (solid) and separate
contributions of ${\cal L}_\s$ (dashed) and  ${\cal L}_4$ (dotted).}
\label{fig1}
\end{figure}

\begin{figure}[hbt]
\centerline{\psfig{figure=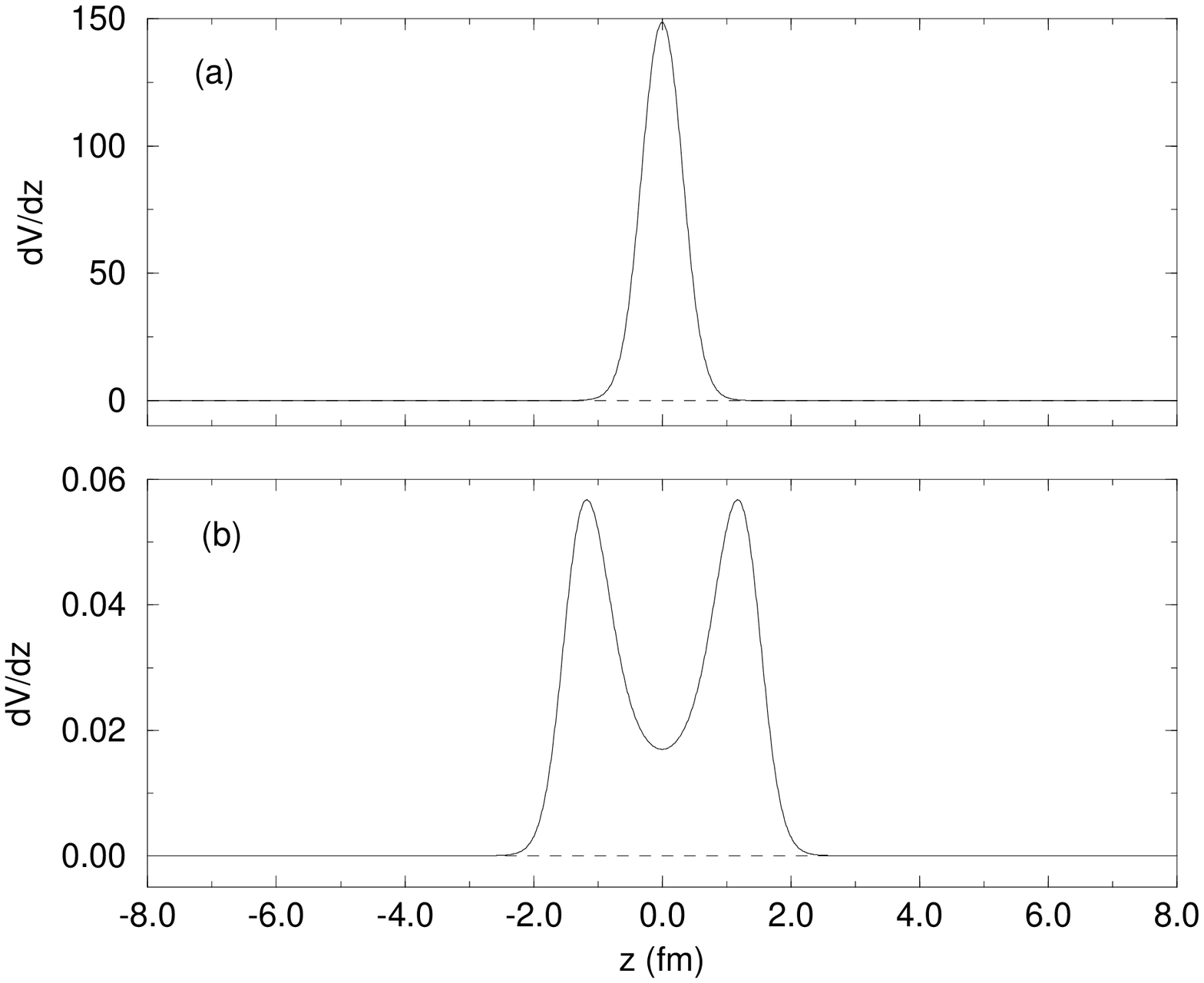,width=6cm,height=10cm}
\psfig{figure=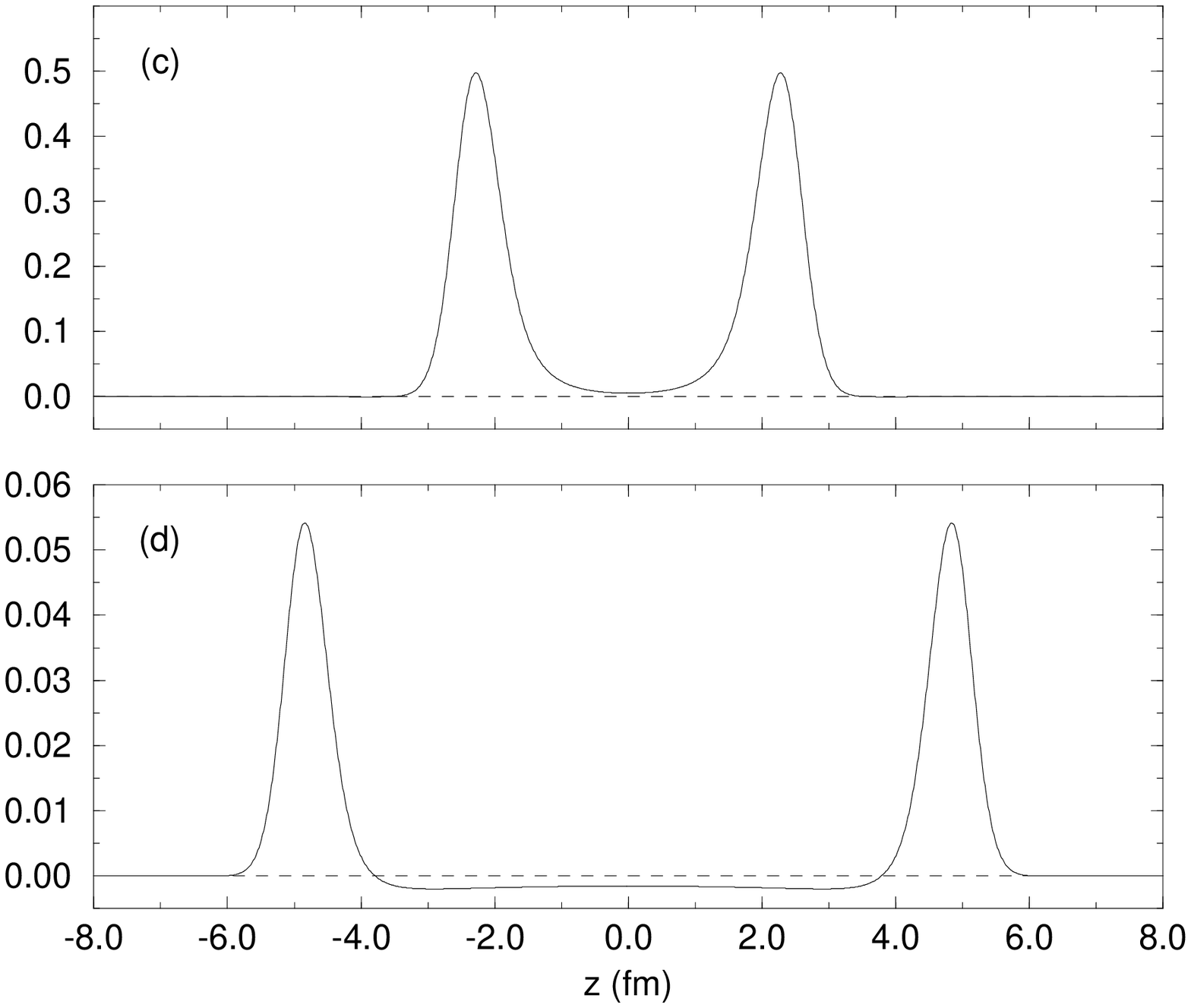,width=6cm,height=10cm}}
\caption{Integrand in $z$ of the scalar potential 
$V_C(d)$, for (a) $d=1$~fm, (b) $d=3$~fm, (c) $d=5$~fm, (d) $d=10$~fm;
vertical axis: 
arbitrary unity.}
\label{cine}
\end{figure}

Inspecting this table, one notes that the part of the potential due to
$\Lg_\s$ is attractive, but is  superseeded by a repulsive
contribution coming from  the stabilizing term.
The net sign of the potential is, then, the outcome of a large cancellation.
On the other hand, the dependence of eq.\ (\ref{asym-form}) on $\id$ is similar to those of both the perturbative chiral calculation,
eq.\ (\ref{va}), and of  the phenomenological Argonne potential.
We stress that this correct geometry is a general feature 
of the model, because it depends only on the form of the $B=1$ solution, 
and not on the specific ansatz used to obtain the $B=2$ result. 
In fig.\  \ref{fig1} 
we display the ratios between the full and asymptotic PA potentials,
as  given by eqs.\ (\ref{vc-form}) 
and (\ref{asym-form}), with the purpose of illustrating their convergence.
The interplay between the attractive contribution from $\Lg_\s$ and
the repulsive one from $\Lg_4$ can also be seen in 
fig.\  \ref{cine}, 
where we present the function $dV_C^{pa}(d)/dz$, corresponding to the integrand in $z$ of expression (\ref{vc-form}), for several values of $d$.
One notes that the contributions in the neighbourhood of the skyrmion centers are large and positive but, on the other hand, 
a negative region develops as the distance $\id$ increases.

These features of the central potential allow us to identify clearly the stabilizing term as the responsible for its repulsive 
character.
Therefore, mechanisms which can reduce the importance of $\Lg_4$ may help in producing an attractive interaction.
In the next section we discuss a class of such mechanisms, associated with deformations of the QCD vacuum.

\section{Constructing ${\bf B=2}$ solutions}
\label{s3}

We consider here $B=2$ solutions, constructed by using the hedgehog $B=1$ skyrmions as building blocks, in the framework 
of the sudden approximation.
In general, an ansatz is a prescription of the form
\beq
U (\bx, \by) = f \lb U (\bx), U (\by) \rb,
\label{pr}
\eeq
where $f$ is a function, chosen according to physical criteria.
The construction of such a function should follow some guidelines:\\
\ni
{\bf 1.}~the baryon number of the composite configuration must be 
two for all distances $\id$;\\
{\bf 2.}~the composite pion field must have the correct quantum numbers, being pseudoscalar, isovector and odd under G-parity;\\
{\bf 3.}~the composite Lagrangian must be chiral symmetric, even under G-parity and invariant under the exchange of the 
two constituent skyrmions.

The standard constructive approximate solution to the $B=2$ system is based on the PA, as discussed in the previous section.
In this approach, the composite pion field, obtained from
eqs.\ (\ref{U}) and (\ref{pa}), is given by   
\beq
\bP_{pa} = \frac{1}{\fpi} \lp \s_\bxm \bpi_\bym + \s_\bym \bpi_\bxm -
\bpi_\bxm \times \bpi_\bym  \rp \,,
\label{pipa}
\eeq
where $\bpi_\brm$ is the pion  field of the hedgehog with coordinate 
$\br$ and $\s_\brm \equiv \fpi \cos F(\br)$.
The function 
\beq
S_{pa} = \frac{1}{\fpi} \lp \s_{\bxm} \s_{\bym} - \bpi_\bxm
\cd \bpi_\bym  \rp\,.
\label{sigpa}
\eeq
is the composite analogous of $\s$ and satisfies $S_{pa}^2 +
\bP_{pa}^2 =\fpi^2$. 

The field $\bP_{pa}$  has a rather serious drawback as a candidate for
the pion field, namely that it contains an azimuthal  
term which is both even under G-parity and antisymmetric under
hedgehog  exchange.
Hence it does not have good pion quantum numbers, violating
requirements 2 and 3 stated above. 

This motivated us to try to understand whether this problem could be
responsible for the absence of  
attraction found in the central potential. We 
considered several alternative possibilities, inspired in the PA.
The basic idea is to propose a composite field $\bP$, use it to define
a function $S$ by  
\beq
S^2 = \fpi^2 - \bP^2 \, , 
\label{esse}
\eeq
construct the unitary field as 
\beq
U = \lb S+i\btau\cd\bP\rb /\fpi \,,
\label{UU}
\eeq
and feed it into the Skyrme Lagrangian.
We begin by describing briefly some unsuccessful attempts, in order to
prevent readers from repeating them.  

The simplest exchange-symmetric ansatz would be the average  
$ \bP =  \lp \bpi_{\bxm} + \bpi_{\bym} \rp / 2$.
However, when $d = 0$, one has  $F_\bxm = F_\bym = F_\brm$ and hence $\bP =
\fpi \sin F_\brm \, \hbr$  corresponds to a $B=1$ field, 
which must be disregarded. 

\begin{figure}[!]
\centerline{\psfig{figure=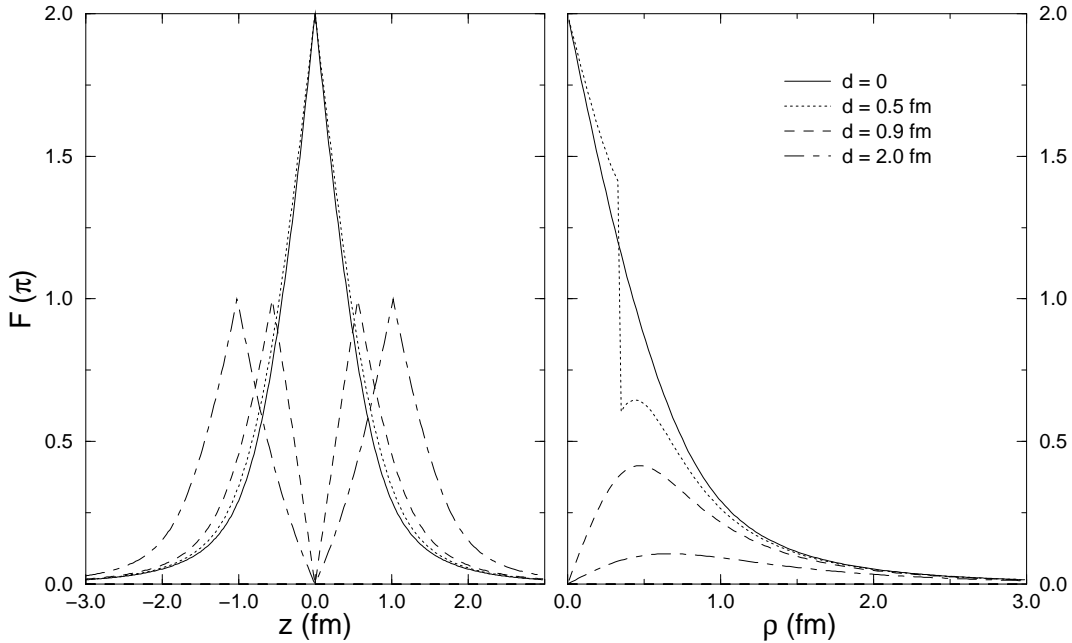,height=9cm}} 
\caption{Profile function $F$ for the product ansatz, in unities of
$\pi$, along the $z$ (left) and $\rho$ (right) axes, for various
values of $\id$.}
\label{perfis}
\end{figure}

This suggests that, in order to obtain $B=2$, it is mandatory to mix $\bpi$ and $\s$. 
In the case of the PA, which yields $B=2$ at all distances, we note that the chiral constraint between $\bP_{pa}$ and $S_{pa}$ 
allows one to write
\beq
U_{pa}= \lb S_{pa} + i \btau\cd \bP_{pa} \rb /\fpi 
\equiv e^{-i \, \btau \, \cd \,\bu \, F_{pa}}\,,
\label{upa}
\eeq
where $\bu$ is a unit vector, taken as pointing always away from the
origin of the coordinate system, and $F_{pa}$ is a  
profile function.
In fig.\  \ref{perfis}  we display the behaviour of this angle along the axes $z$
and  $\r$, for various 
values of the internucleon distance $\id$. 
The solid lines correspond to
the case $\id=0$, which is 
spherically symmetric and it is possible to see that, along both
directions, the  chiral angle varies smoothly from $2\p$ at the 
origin to $0$ at infinity.
In the case $\id=0.5$ fm, shown in dotted lines, 
one notes that a discontinuity has appeared along 
the $\r$ axis.
This discontinuity increases with distance and, at $\id_{crit}=0.86$
fm,  the chiral angle is such that 
$F_{pa}(\r=0, z\rar 0) =2\p$ and $F_{pa}(\r\rar 0, z= 0) = 0$.
Therefore, at this critical point, it is more natural to set 
$F_{pa}(0,0)=0$ and to work with two separate
solutions, 
such as
illustrated by the dashed and dot-dashed lines, corresponding to $0.9$
and $2.0$ fm, respectively.
This suggests that, from $\id_{crit}$ onwards, each of the interacting
skyrmions acquires a considerable  individuality.

\begin{figure}[!]
\centerline{\psfig{figure=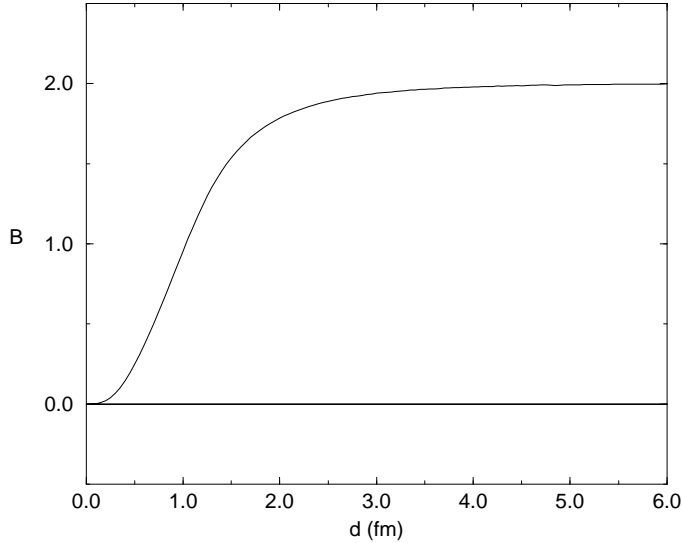,width=9cm}} 
\caption{Baryon number as a function of distance $d$ (fm), for the
ansatz given by eqs.\  (\ref{nov}) and (\ref{esse}).} 
\label{nb_errado}
\end{figure}

The combination 
\beq
\bP = \frac{1}{f_\pi} \lp \s_{\bxm} \bpi_{\bym} + \s_{\bym}
\bpi_{\bxm} \rp 
\label{nov}
\eeq
is interesting, for it has an explicit physical meaning.
As the function $\s(\br)$ is associated with the quark condensate that
surrounds the baryon labeled by $\br$, 
this field $\bP$ represents each skyrmion immersed in the distorted vacuum of the other one.
The condition (\ref{esse})  allows one to determine $S$ up to a sign.
In the case $B=1$ the field $\s$ changes sign when one goes from infinity to the origin and the same happens when $B=2$.
The sign of $S$ is also important and, in order to fix it, we note that
the behaviours of eqs.\ (\ref{pipa}) and (\ref{nov}) along the $z$ axis
are identical, 
since the azimuthal component vanishes. 
We then forced the condition $S=S_{pa}$ along this axis.
However, this ansatz, based on eq.\ (\ref{nov}), gives rise to a baryon
number which varies with $\id$, as shown in fig.\  \ref{nb_errado}, 
and had to be abandoned.

\begin{figure}[!]
\centerline{\psfig{figure=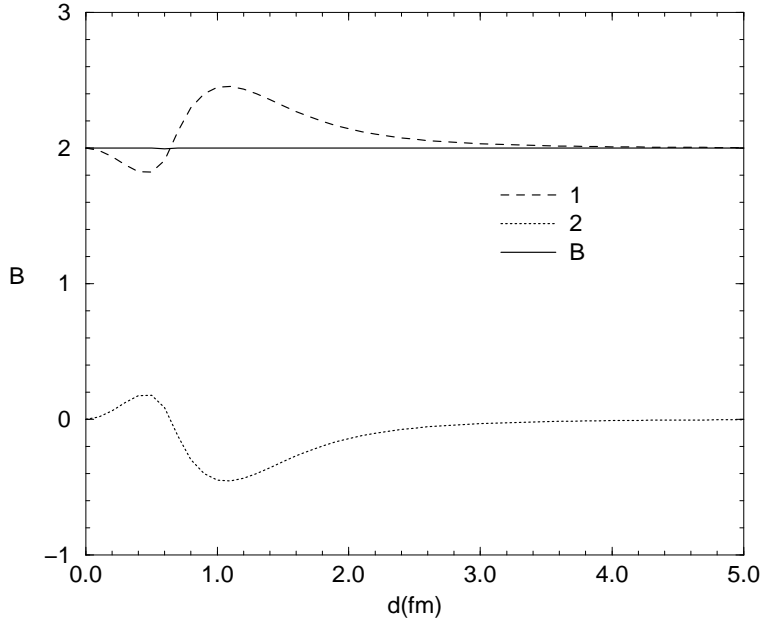,width=10cm}} 
\caption{Baryon number as a function of the separation distance (fm)
for the symmetrized ansatz (solid); separate contributions $B_1$
(dashed) and $B_2$ (dotted),  as defined
in eq.\  \ref{bzerox}.}
\label{nb}
\end{figure}

This discussion illustrates the fact that it is not trivial to build
an ansatz with a good topology.
We thus decided to adopt simultaneously the pion field as given by
eq.\  (\ref{nov}) and the function $S_{ap}$ of the PA,
eq.\  (\ref{sigpa}), for its topology is automatically correct. 
With this option, the unitarity constraint reads
\beq
S_{ap}^2 + \bP^2 = \fpi^2\; \rad^2,
\label{vinc_na}
\eeq
where
\beq
\rad = \sqrt{1 + \lb \lp \bpi_{\bxm}\cd \bpi_{\bym} \rp^2 - \bpi^2_{\bxm}\; \bpi^2_{\bym} \rb / \fpi^4 } 
\label{eta_na}
\eeq
and the pion field becomes in fact $\bP/\rad$.
This form for the dynamical variable is the same as that proposed by
Nyman and Riska, in their symmetrized product ansatz \cite{nr88}.  
This ansatz has a topology similar to  the PA, as illustrated in 
fig.\ \ref{perfis}.
The corresponding baryon number density is given in appendix A and, 
in the classical case,
yields $B=2$ for all distances when integrated over space, as shown in
fig.\  \ref{nb}. 

\section{Results and conclusions}
\label{s4}

In order to derive the potential, we use the quantized fields of
eq.\  (\ref{q}), obtained by rotating the constituent skyrmions. 
This idea of rotating individual hedgehogs corresponds to an
approximation and deserves some attention.  
The quantization of a hedgehog, as discussed in sect. \ref{s2},  amounts to
multiplying the classical field by the  
matrix $D$, which depends on three free parameters.
In the case $B=1$, this procedure does not change the baryon current.
This can be seen by writing the baryon density for quantized fields as 
\beq
B^0  =  -\;\frac{1}{12\p^2}\;\e_{abc}\;\e_{\a\b\g}\;\frac{1}{\s}\;
\dr_a D_{\a i}\p_i \; \dr_b D_{\b j}\p_j \; \dr_c D_{\g k}\p_k 
\label{bdq}
\eeq
and using the result
\beq
D_{\a a} D_{\b b} = \frac{1}{3}\;\d_{\a\b} \d_{a b} + \frac{1}{2} \;\e_{\a\b\g} \;D_{\g c}\;\e_{cab}
+ \, \mbox{isotensors} 
\label{DD}
\eeq
in order to obtain
\beq
B^0  =  -\;\frac{1}{12\p^2}\;\e_{ijk}\;\e_{abc}\;\frac{1}{\s}\;
\dr_a \p_i \; \dr_b \p_j \; \dr_c \p_k \;.
\label{bdc}
\eeq
This shows that the baryon density is the same for both quantized and
classical fields. 

Analogously, in the case $B=2$, quantization would require a matrix $\bar{D}$, depending on six collective coordinates.
However, the determination of this general matrix may prove to be very
difficult and, in the spirit of  the 
the sudden approximation,  one normally uses $\bar{D}\approx
D^{(\bxm)} I^{(\bym)}  + I^{(\bxm)} D^{(\bym)}$, where $I$ is an
identity matrix and 
$D^{(\bxm)} , D^{(\bym)}$
are operators over the skyrmions labeled by $\bx$ and $\by$ respectively.
The price one pays for this approximation is that it leads to a
quantized baryon number which depends on $D^{(\bxm)}$ and
$D^{(\bym)}$.
This happens because the relation equivalent to eq.\ (\ref{DD}) does not
hold for the approximate matrix $\bar{D}$ 
and hence does not represent a major shortcoming for the symmetrized
ansatz (SA).
Indeed, as pointed out by 
Sternheim and K\"albermann \cite{sk89}, this poses problems for short
distances only.

A collective rotation of the pion field $\bP$, as in the $B=1$ case, would leave $S/ \rad$ unmodified, as a classical function. 
However, this would also mean to treat the scalar product $\bpi_\bxm
\cd \bpi_\bym$ as a classical quantity and would lead to serious
contradictions, for the OPEP content of the isospin dependent channels
relies on the quantum character of such a scalar product  
in eq.\ (\ref{sigpa}).

Therefore, the individual rotation of each pion field is more consistent with a constructive approach, 
although not free of problems.
In principle, every pion field $\bpi$ in the composite Skyrme
Lagrangian should be quantized. 
When applying this prescription to the dynamical variable of the
SA, one has to deal with the functions $\rad^{-2}$ and
$\rad^{-4}$,  
which depend on pion fields, coupled to operators $D$.
The meaning of the quantized $\rad^{-2}$ is that of a power series in
$D$, which involves products of arbitrary  
numbers of these matrices and hence can only be handled by resorting
to truncation. 
With this limitation in mind, we treat $\rad^{-2}$ as a polynomial in
$\bpi$ and thus its expectation value between two-nucleon states can
be evaluated without ambiguities. 

In order to test the implications of this assumption, in the sequence we present results with two versions of 
$\rad$, namely, a classical one, 
\beq
\rad_c^{-2} =  \lc 1 - \lb 1- (\hbx \cd \hby )^2 \rb s_\bxm^2 s_\bym^2
\;\rc^{-1} ,
\label{eta-clas}
\eeq
and a quantized one, truncated at the first order in the $D$
expansion, given by 
\bea
\la NN| \rad_q^{-2} |NN\ra &=& \la N N | \lc 1 - s_\bxm^2 s_\bym^2 
+ D^{(\bxm)}_{\a i} D^{(\bym)}_{\a j} D^{(\bxm)}_{\b k} D^{(\bym)}_{\b \ell}\; 
\lp \pi_{\bxm} \rp_i \lp \pi_{\bym}\rp_j   \lp \pi_{\bxm}\rp_k \lp
\pi_{\bym} \rp_\ell  / \fpi^4 \;\rc^{-1}
| N N \ra  
\nonumber \\[2mm]
&\approx&  \lc 1 - \frac{2}{3}\; s_\bxm^2 s_\bym^2 \rc^{-1} .
\label{eta-q}
\eea

Replacing the pion field of the symmetrized ansatz into the interaction Lagrangian used to calculate the potential,
one has 
\beq
\Lg_{int} = \Lg^S +  D^{(\bxm)}_{\a m} D^{(\bym)}_{\a n} \; \Lg_{mn}^V \;, 
\label{eq452a}
\eeq
where the labels $S$ and $V$ stand respectively for isoscalar and isospin dependent parts of $\Lg_{int}$.
Using this result in eq.\ (\ref{pot1}), one gets 
\beq
V(\id)=V_C + \btau^{(\bxm)} \cd \btau^{(\bym)} \lb \bsig^{(\bxm)} \cd
\bsig^{(\bym)} V_{\!S\!S}+  S_{12}  V_T  \rb \;, 
\label{pot2}
\eeq
where $V_C$, $V_{\!S\!S}$ and $V_T$ are the usual central, spin-spin
and tensor components. 
All terms receive both G-parity odd and even contributions.

\begin{figure}[htb]
\centerline{\psfig{figure=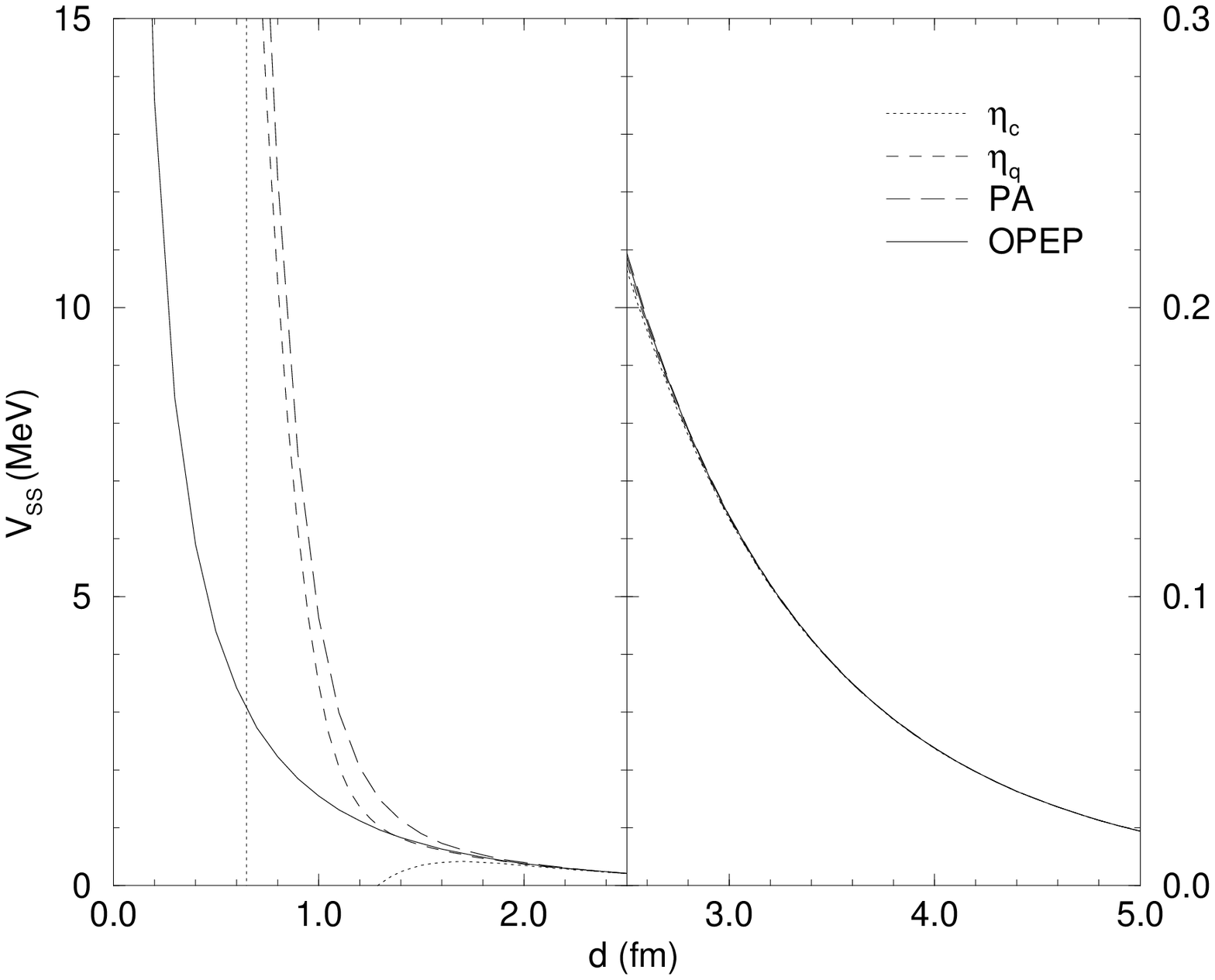,height=7cm}
\psfig{figure=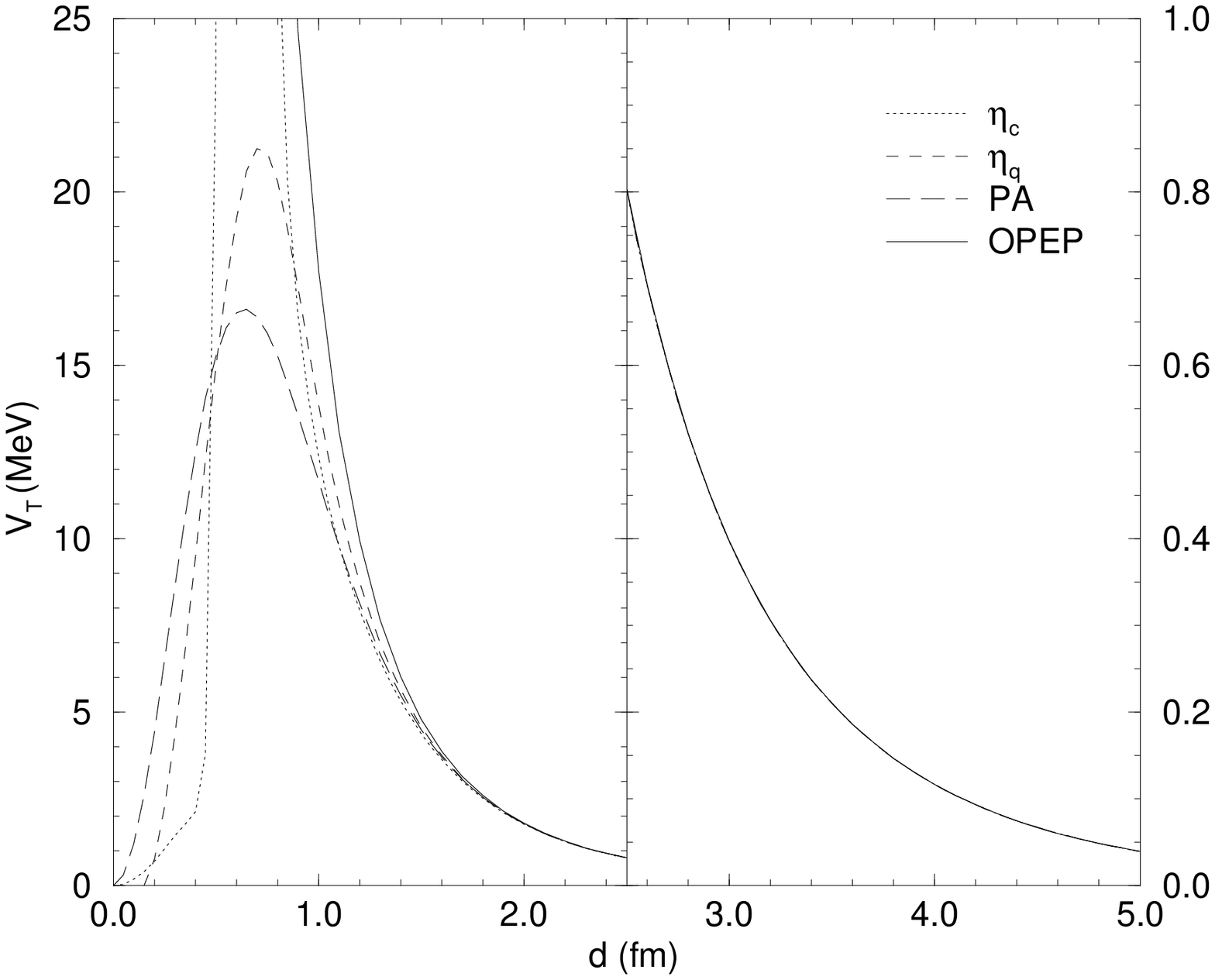,height=7cm}}
\caption{Comparison among  
Skyrme model G-parity odd spin-spin (left) and 
tensor (right) potentials 
from symmetrized ansatz, with classical
(dotted) and quantized (dashed) versions of $\rad$, from product
ansatz (long dashed) and
OPEP (solid).}
\label{opep}
\end{figure}

\begin{figure}[!htb]
\centerline{\psfig{figure=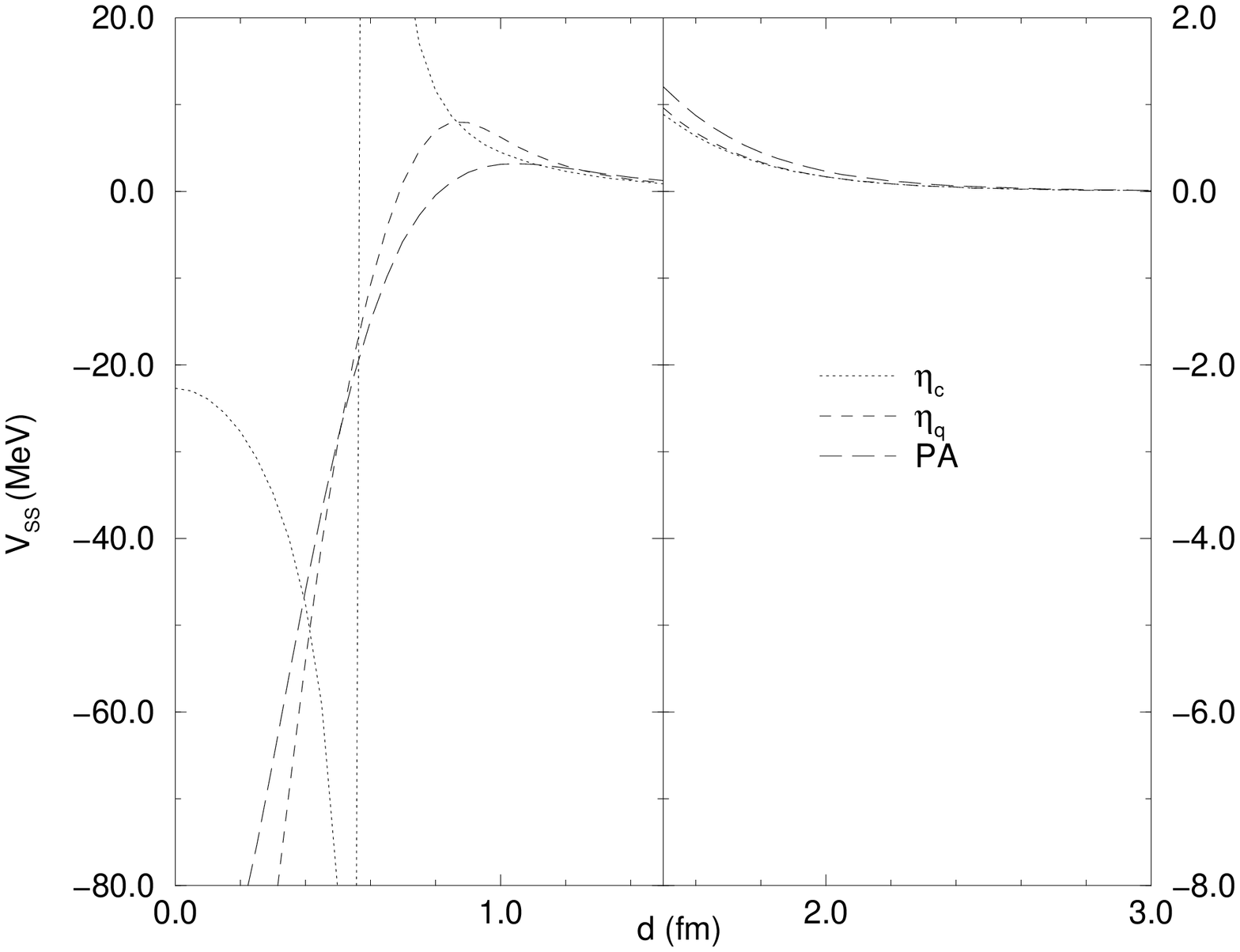,height=6.5cm}
\psfig{figure=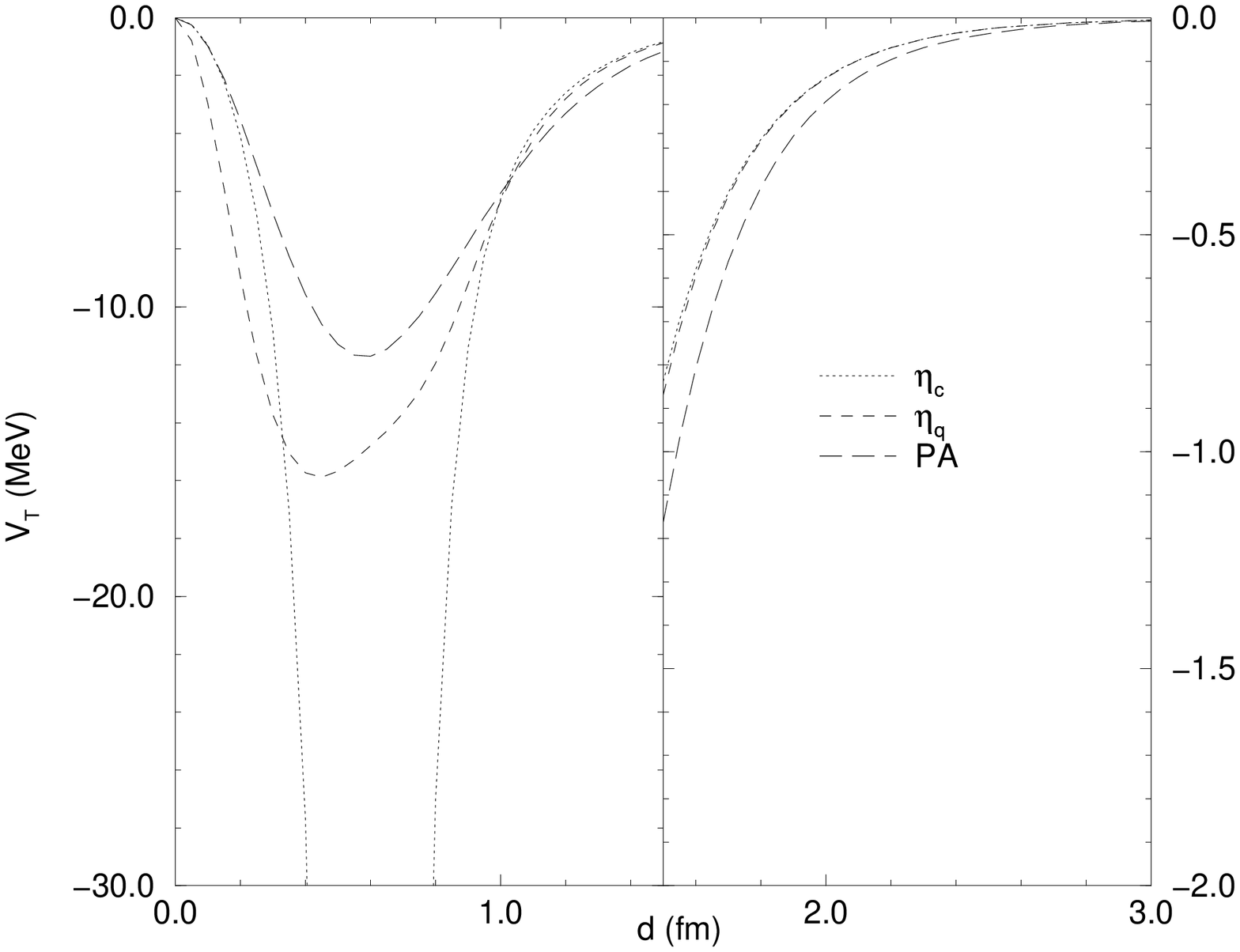,height=6.5cm}
} 
\caption{G-parity even spin-spin (left) and tensor (right) 
potentials from  symmetrized ansatz, with 
classical (dotted) and quantized (dashed) versions of 
$\rad$, and from product ansatz (long dashed).}
\label{pp}
\end{figure}

The G-parity odd components of spin-spin and tensor terms of the
potential are shown in fig.\  \ref{opep}, for the
two possible choices of $\rad$, compared to the PA and pure OPEP results. 
One sees that all curves coincide for distances larger than 2 fm,
indicating that all ans\"atze  reproduce  asymptotically
the OPEP. 
The results for the G-parity even terms are of minor importance here,
as we are interested in the long range  behaviour 
of the potential, but they are included for completeness in
fig.\  \ref{pp}. One should note that in the case of $\rad_c$, the
potentials present a singularity at $d \sim 0.6$ fm,
due to a root of $\rad_c$. 

\begin{figure}[!htb]
\centerline{\psfig{figure=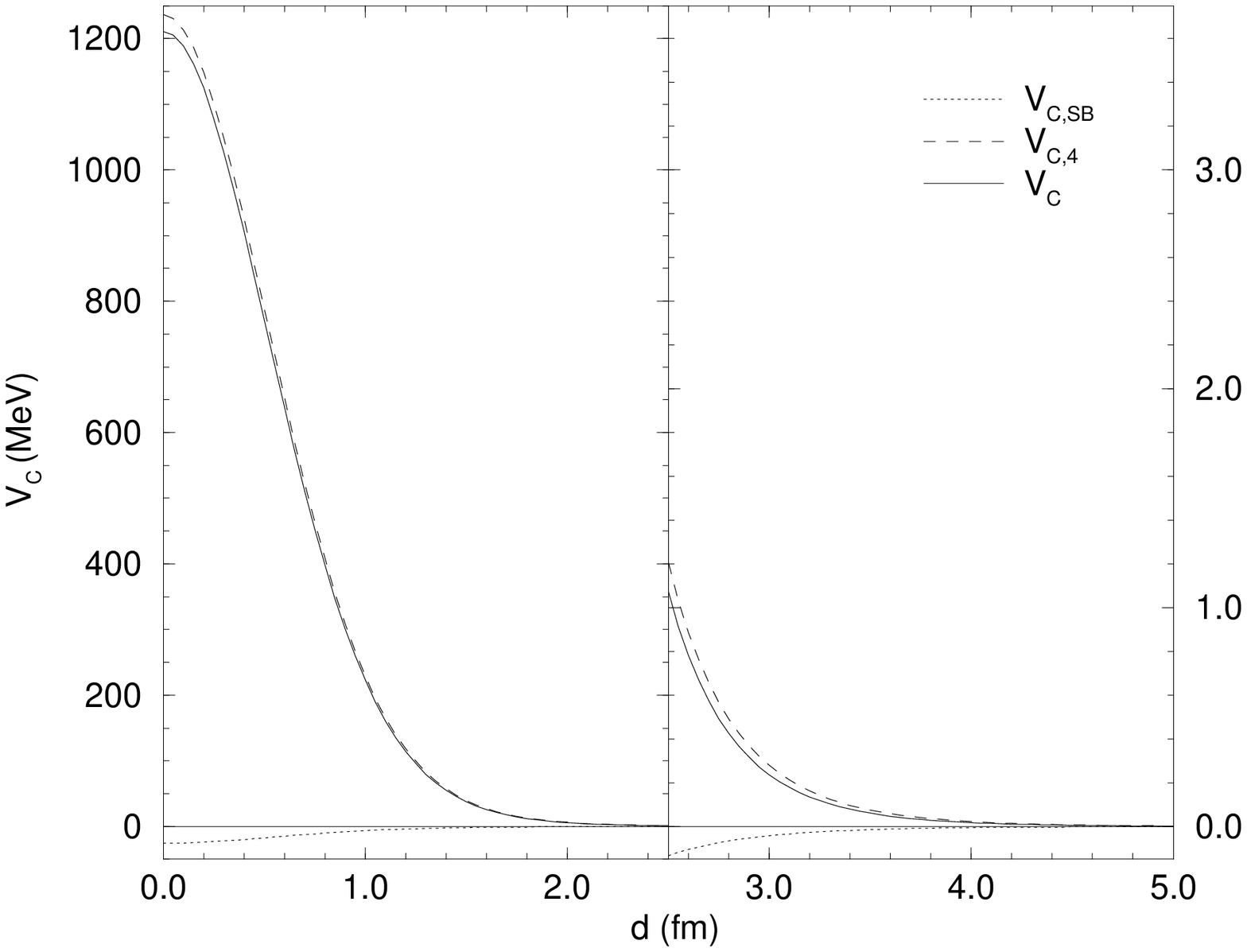,height=7cm}
\psfig{figure=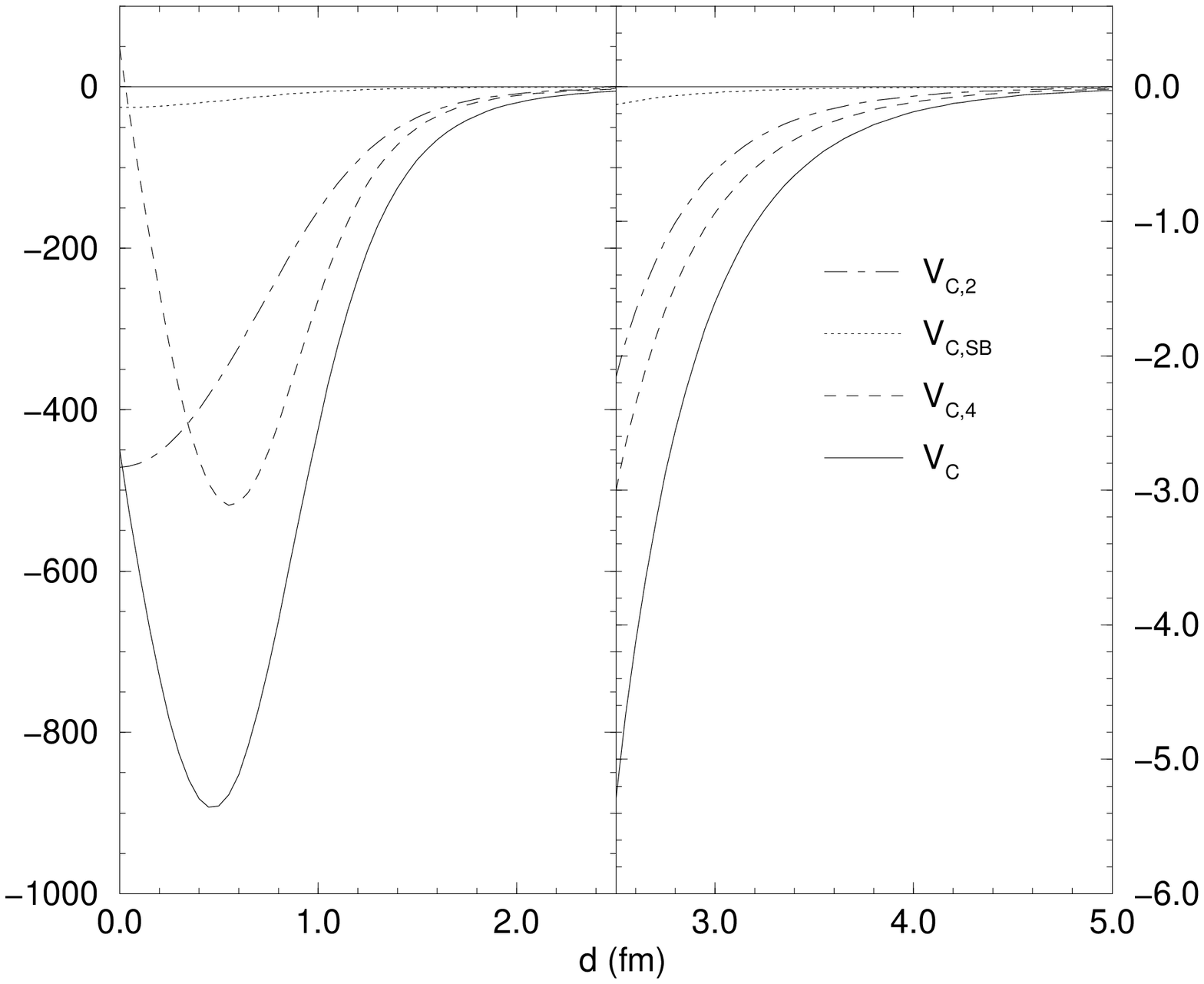,height=7.0cm}
}
\caption{Scalar potential in the Skyrme model: total result and
components. 
Product ansatz (left) and symmetrized ansatz with $\rad = 1$ (right).
} 
\label{vc1}
\end{figure}

\begin{figure}[!htb]
\centerline{\psfig{figure=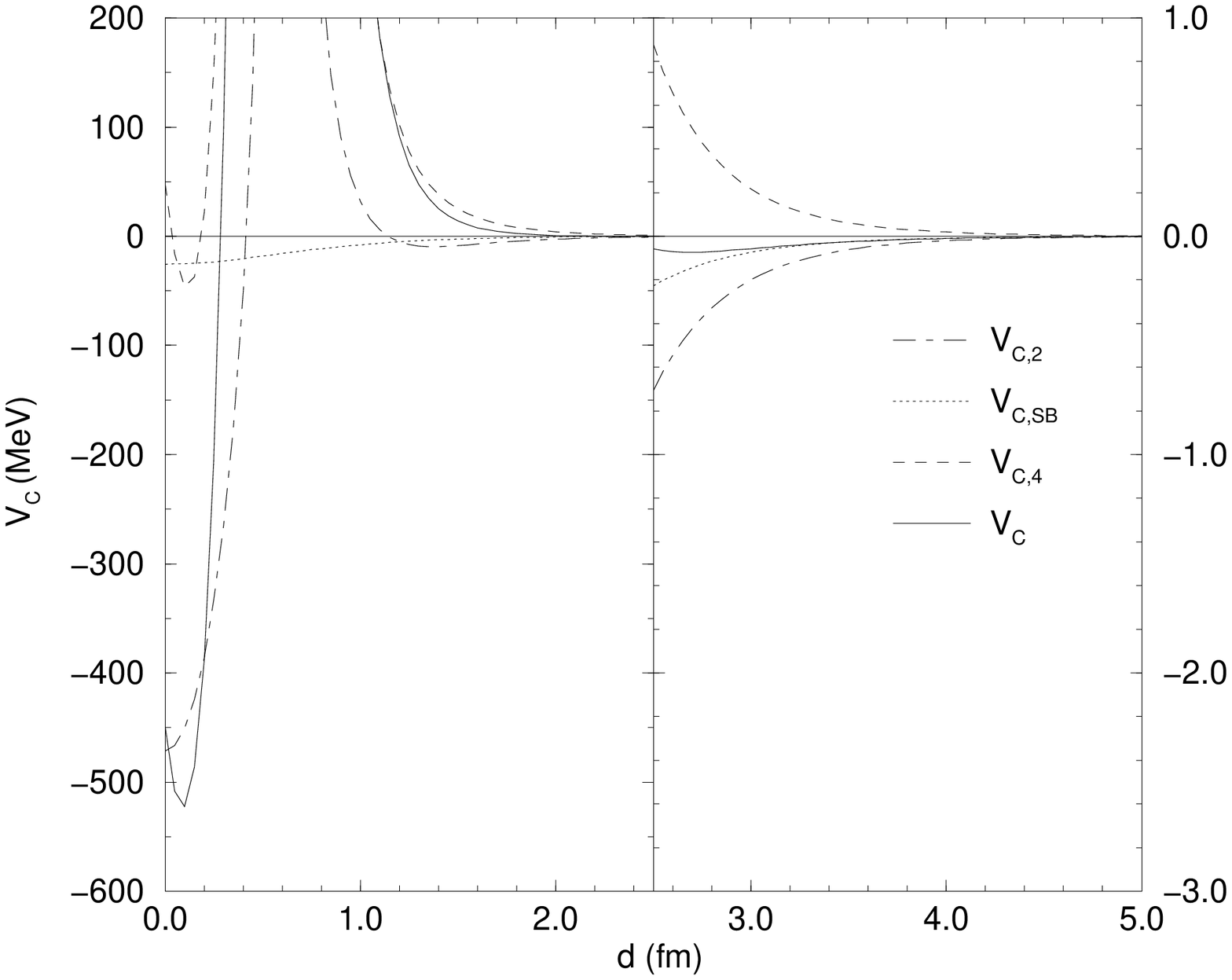,height=7.5cm}
\psfig{figure=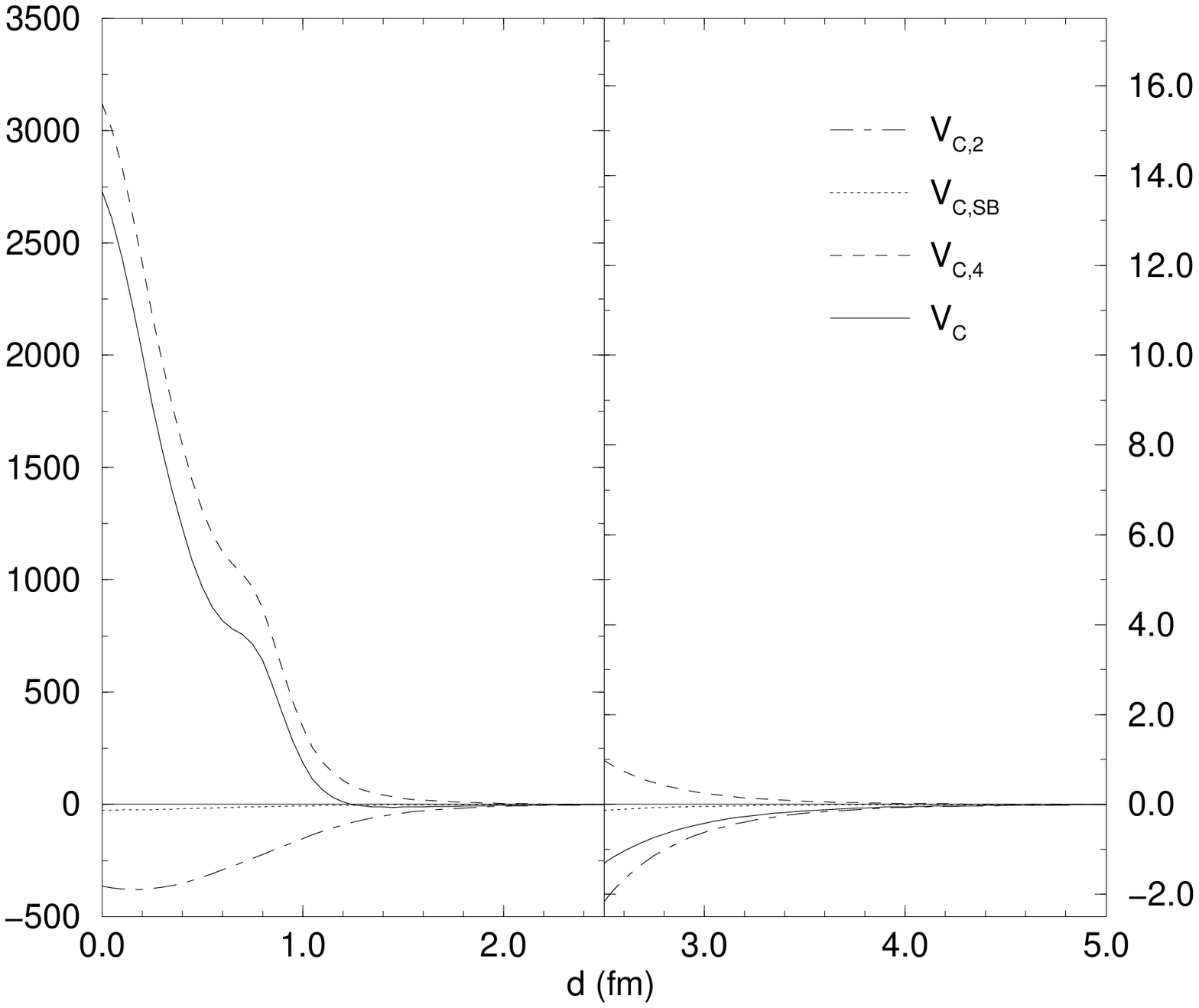,height=7.3cm}
}
\caption{Scalar potential in the Skyrme model with the symmetrized
ansatz: total result and
components; 
$\rad_c$ (left) and $\rad_q$ (right).
} 
\label{vc2}
\end{figure}

Results for the scalar component $V_C$ are presented in
figs.\ \ref{vc1} and \ref{vc2}. In the former we display the behaviour 
of  the PA
(left) and the predictions from the SA with $\rad = 1$ (right), 
which is non-unitary and considered just for pedagogical purposes.
Inspecting it one learns that the SA includes
a contribution from $\Lg_2$, that was not present in the PA. Moreover the 
contribution from
$\Lg_4$ is negative, and so is the net result for $V_C$.

The two valid options for the SA considered here, based on 
$\rad_c$ and $\rad_q$, are given in fig.\  \ref{vc2}.
In both cases we observe that the unitarity constraint restores the
repulsion due to $\Lg_4$, but in such a way that the net
result is asymptotically attractive. 
On the other hand, the amount of overall attraction found in the SA
depends on the specific quantization prescription adopted. 
At very large distances, the curves corresponding to $\rad_c$ and
$\rad_q$ have the same geometry 
and yield respectively the following approximate values for the 
intensity of the potential: $K_c = 14$ MeV and $K_q = 57$ MeV. 
Comparing them with the empirical values in table \ref{t1}, it is
possible to see that predictions from the SA 
are qualitatively reasonable.

In summary, we have shown that the SA provides the correct baryon
number for the two-nucleon system in the Skyrme model, as well as 
 attractive central $NN$ potentials. The 
correct quantitative feature is somewhere between the values obtained
for the two versions of the normalization function $\rad$. 

We conclude that it is indeed relevant to the central potential to
eliminate the term with a wrong G-parity from $\bP_{pa}$.
We expect that a deeper and more careful study of the
quantization procedure will lead to a more accurate evaluation of the
amount of attraction coming from the SA.

\section*{Acknowledgements}
The work by I.P.C. was supported by Funda\c c\~ao de
Amparo \`a Pesquisa do Estado de S\~ao Paulo - FAPESP (Brazilian
agency), grant 94/1801-4. We would like to thank the
hospitality of the Nuclear Theory Group, in the University of
Washington, and of the Division de Physique Th\'eorique, in the
Institute de Physique Nucl\'eaire/Orsay, where part of this work was
performed.

\appendix
\section{Appendix: The baryon number}

The explicit calculation of the zero-component of the baryon current
yields 
\beq
B^0 = B^0_1 + B^0_2 \,, 
\label{bzerox}
\eeq
where
\begin{eqnarray*}
& & B^0_1 (\br ; \id) = - \frac 1 {2 \pi^2 \, {\rad}^2 } 
\left(c_\bxm {\frac{s_\bym}{\bym}} + c_\bym {\frac{s_\bxm}{\bxm}}  
\right)^2 \left(F_\bxm^\prime +
F_\bym^\prime \right) \, ,  \hspace*{8cm} 
\\[2mm]
& & B^0_2 (\br ; \id) = - \frac 1 {2 \pi^2} \frac 1 { 
\left( c_\bxm c_\bym - \left(\hbx \cdot \hby\right) 
s_\bxm s_\bym \right) {\rad}^2 } 
\left(c_\bxm {\frac{s_\bym}{\bym}} + c_\bym {\frac{s_\bxm}{\bxm}} \right)
\\[2mm]
\lefteqn{ \cdot \left\{ 
\left(1-\left(\hbx \cdot \hby\right)^2\right) 
\left[ c_\bxm
c_\bym \left( c_\bxm F_\bxm^\prime - {\frac{s_\bxm}{\bxm}}\right) 
\left( c_\bym F_\bym^\prime -
{\frac{s_\bym}{\bym}}\right) - s_\bxm^2 s_\bym^2   F_\bxm^\prime  
F_\bym^\prime \right]  
\right.} 
\nonumber 
\\[2mm]
\lefteqn{ 
- \left. \left[ \frac {s_\bxm s_\bym}{ {\rad}^2} J_\bxm 
\left( \left(c_\bxm {\frac{s_\bym}{\bym}} + c_\bym {\frac{s_\bxm}{\bxm}} \right) 
\left(c_\bym s_\bxm + \left(\hbx \cdot \hby\right) c_\bxm s_\bym \right) 
+ \left(1-\left(\hbx \cdot \hby\right)^2\right) s_\bxm c_\bxm 
\left(  F_\bym^\prime - c_\bym {\frac{s_\bym}{\bym}} \right) \right)
\right] 
- \left[ \bxm \leftrightarrow \bym \right] 
\right\}\, , 
}  
\nonumber 
\end{eqnarray*}
with
$
J_\bxm = \left(\left(\hbx \cdot \hby\right)^2 -1 \right) 
c_\bxm F_\bxm^\prime s_\bym + \left(\hbx \cdot \hby\right) 
\left(
s_\bxm {\frac{s_\bym}{\bym}} - \left(\hbx \cdot \hby\right) 
s_\bym {\frac{s_\bxm}{\bxm}} \right)$.

The numerical integration of $B^0_2$ is tricky due to the presence of
the function $S$ in the denominator. Results are
shown in fig.\  \ref{nb}, as functions of separation distance
$d$. It shows that the SA presents the correct topology for the
$NN$ system.


\end{document}